\documentstyle[epsfig]{mn}

\title[The UVX quasar luminosity function and its evolution]
{\noindent{\huge\bf 
The UVX quasar optical luminosity function and its evolution}}

\author[P. Goldschmidt \& L. Miller]
{{\large Pippa Goldschmidt$^1$ and Lance Miller$^2$}\vspace*{0.1in}
\\
{$^1$Astrophysics Group, I.C.S.T.M., Prince Consort Rd., London SW7 2BZ, U.K.}\\
{$^2$Dept. of Physics, Oxford University, Keble Road, Oxford OX1 3RH, U.K.}}

\begin{document}
\maketitle
\begin{abstract}
The recently-finished Edinburgh UVX quasar survey at $B < 18$
is used together with other complete samples to estimate the shape
and evolution of the optical luminosity 
function in the redshift range $0.3 < z < 2.2$.  There is
a significantly higher space density of quasars at high luminosity and low
redshift than previously found in the PG sample of Schmidt \& Green (1983),
with the result that the shape of the luminosity function at low redshifts 
($z < 1$) is seen to be consistent with a single power-law. 
At higher redshifts the slope of the
power-law at high luminosities appears to steepen significantly.  
There does not appear to be any consistent break feature 
which could be used as a tracer of luminosity evolution in the population.
\end{abstract}

\section{INTRODUCTION}

One of the strongest pieces of evidence for an evolving universe has long been 
the observed evolution in comoving space density of the quasar population (Schmidt 1968).  
Until recently it had been thought that the shape of the quasar 
optical luminosity function and 
its evolution over the redshift range $0 < z < 2$ was well understood, 
and attempts to explain the physical causes of quasar evolution have relied 
on attempting to predict the observed evolution of the luminosity function 
(e.g. Haehnelt \& Rees 1993).
However recent studies (Goldschmidt et al., 1992, Hewett et al., 
1993, Hawkins \& V\'{e}ron 1993 \& 1995) have cast doubt 
upon the completeness of the surveys used to define the luminosity function. 
In this paper we present an analysis of the quasar luminosity function and 
its evolution based on new observational data,  and argue that knowledge of 
the luminosity function alone is insufficient to allow us to understand the 
physical causes of quasar evolution.

The most widely-quoted study of the luminosity function to date has been 
that of Boyle et al. (1988, hereafter BSP), who used the AAT sample of 
faint UVX quasars
together with brighter samples such as the Palomar-Green survey (Schmidt \&
Green, 1983) to determine the luminosity function in four redshift slices from
$z=0.3$ to $z=2.2$. 
BSP fit a variety of models to this function and concluded that the best-fit
model was pure luminosity evolution (PLE) in which the shape of the luminosity 
function was parameterised by two power-laws with a transition between them 
at a characteristic luminosity.
The characteristic luminosity increased with redshift (we term this negative evolution,
as the population appears to have become dimmer with increasing cosmic time):
\begin{eqnarray}
\frac{d\phi}{dM}(M,z) = \frac{\phi^*}{\left[10^{0.4(\alpha+1)(M-M^{*}(z))} 
 +10^{0.4(\beta+1)(M-M^{*}(z))}\right]}
\end{eqnarray}
in which $\alpha$ and $\beta$ are the indices of the power-laws and $M^{*}(z)$
describes the evolution;
\begin{equation}
	M^{*}(z) = M^{*}_{0} - 2.5k\log_{10}(1+z)
\end{equation}
With the addition of two surveys extending to redshift $z < 2.9$ (Boyle,  Jones
\& Shanks 1991, Zitelli et al. 1992) 
the above model has been slightly modified such that there is a maximum
redshift beyond which no evolution occurs. The most recent parameters of this
model have been presented by Boyle (1991) and are; 
$\alpha=-3.9$, $\beta=-1.5$, $k=3.5$, $M^{*}_{0}=-22.4$,  $z_{max}=1.9$. 

BSP ruled out any need for additional density evolution for quasars with $M_{B}
\le -23$.
PLE can be interpreted
as either representing the {\it actual} evolution of individual objects in
which a single population of quasars formed at a single epoch and have been
growing dimmer ever since, or as the {\it statistical} evolution of the
properties of successive populations. BSP noted that the latter interpretation
implies a conspiracy between birth and death rates.
However the former interpretation, in which lifetimes are of the order
of the Hubble time, predicts massive remnant black holes in Seyfert
galaxies at low redshift (Cavaliere \& Padovani 1989). 

The analysis of BSP relied on brighter quasar surveys, principally the
Palomar-Green survey (Schmidt \& Green 1983), in order to determine the
most luminous part of the luminosity function. However, doubts have been raised
about the completeness of this survey (Wampler \& Ponz 1985) and initial
results from the Edinburgh Multicolour Survey (Goldschmidt et al. 1992) showed
that the Palomar-Green survey under-estimated by a factor 3 the surface
density of quasars with $B \le 16.5$.  In
this paper we shall replace the Palomar-Green data with data 
from the Edinburgh survey.

Hewett et al. (1993) presented the first estimates of the space density of
quasars from the recently completed LBQS (Morris et al. 1991 and references
therein), and compared those estimates with those predicted by the PLE model.
They found that the PLE model over-predicts the number of quasars at the faint
end and under-predicts the number of quasars at the luminous end
of the luminosity function in each redshift slice. They show that the slope of
the luminosity function for luminous quasars appears to change shape with redshift, in
contradiction to PLE, and argue that modification of the model is needed:  
we shall see that our new results are in accord with that conclusion. 

Hawkins \& V\'{e}ron (1993 \& 1995) have used variability-selected 
samples of quasars to calculate the luminosity function in the same 
flux range as the AAT survey and conclude that the latter
survey is incomplete, and that the characteristic luminosity, or ``break'', 
detected by BSP is simply an artefact of this incompleteness.  
If the luminosity function is a single power-law with no break then
there is no way of discriminating between luminosity and density evolution. 

This paper presents the results from the recently finished Edinburgh
Multicolour Survey. A brief summary of this survey is given in section 2, 
and in section 3 we use this survey together with fainter UVX surveys to estimate the 
luminosity function in redshift slices. In section 4 we test whether the data can be
adequately described by either the BSP model or even by any 
evolving power-law model in which the power-law index remains constant: 
a class of models which includes those of BSP and Hawkins \& V\'{e}ron (1995).
Section 5 presents the results of fitting empirical models to the data.

\section{THE EDINBURGH QUASAR SURVEY}

A full description of the construction of the survey is given by 
Mitchell (1989), Goldschmidt (1993) and Miller et al. 
(1997), what follows is a brief summary.

The survey is based on 130 U.K. Schmidt telescope (UKST) plates taken in 13
contiguous fields in five wavebands (the photographic bands $u$, $b$, $v$, $r$ and $i$) 
at high Galactic latitude covering 330 deg$^2$.
The coordinates of the field centres range from
$12^h40^m$ to $14^h20^m$ (equinox 1950) in RA at Dec. $-5^{\circ}$ (UKST fields
789 to 794) and from $12^h40^m$ to $14^h40^m$ at Dec. $0^{\circ}$ (UKST fields
861 to 867).  
The plates in each band were taken close
together in time so that incompleteness and contamination due to
variability should be insignificant. 

The plates in each waveband 
were scanned and measured
on the COSMOS machine (MacGillivray \& Stobie 1984) and only those objects
detected on both plates were included in the final dataset.
The resulting                                  
dataset was  calibrated with photoelectric and CCD sequences
in every waveband in every UKST field (Mitchell 1989 \& Goldschmidt 1993), 
obtained at the ESO-Danish 1.5 m., University of Hawaii $88''$, Steward Observatory
$60''$ \& $90''$ and JKT 1 m. telescopes. 

We then used this calibrated dataset to select UVX candidates.
The prime selection criterion for the UVX quasar sample
was $u-b$ colour, requiring quasar  candidates to have $u-b < -0.30$ on
average, although the exact value varied slightly from field to field
(see Miller et al. 1997).

A morphological criterion was also imposed to exclude any candidates 
which appeared extended on
the UKST $u$ plates; the prime reason for this was to exclude blended objects
with peculiar colours which would contaminate the candidate lists. 
Spectroscopic confirmation of all the candidates has been carried
out at the INT 2.5 m., the ESO 1.5 m. and 2.2 m. telescopes  and the UKST,
the latter using the FLAIR multifibre spectrograph.

Spectra have been obtained for a total of 206 quasars, of which 120 with $0.3
\le z \le 2.2$ and $15 \le b \le 18$ form the complete sample used in this paper.  For 
the analysis presented here we 
transform from the
photographic $b$ band to the standard Johnson $B$ band using an average
correction of 0.06 magnitudes.  This was derived from the
transformation of Blair \& Gilmore (1982), $B = b + 0.34(b-v)$, and
assuming the average $b-v = 0.18$ for the quasars in the Edinburgh
survey.

The resulting quasar sample should be complete in the 
redshift and magnitude ranges quoted above. 
The lower redshift limit arises because low
redshift quasars may have host galaxies that are visible and hence appear
extended, or they may have redder colours  due to the underlying host galaxy.
However this limit is poorly determined and is obviously a function of quasar
luminosity. 

\section{THE DIFFERENTIAL LUMINOSITY FUNCTION}

We use the 120 quasars in the complete sample from the Edinburgh survey
together with the AAT sample (Boyle et al.
1990), the SA94 sample (La Franca et al. 1992) and the MBQS sample 
(Mitchell et al. 1984) to estimate the luminosity function. 

We transform the photographic magnitudes in the AAT survey to the standard
Johnson system $B$ using the empirically determined relation in BSP, 
$B = b_{AAT} - 0.1$.
This is not the same as the transformation used for the Edinburgh survey
because of the non-standard photographic magnitude system used for the AAT
survey.

The luminosity function can be estimated using the sum of
the inverse of the comoving volume of the universe searched to find each
object in the survey (Schmidt 1968), where
the available volume is calculated from the minimum and maximum redshifts 
at which an object could have been detected by a given survey,
given its luminosity and the flux limit of the survey 
(note, however, that this method assumes that
locally the comoving space density of quasars is uniform, and hence, 
if binned over large redshift ranges,
strong evolution 
leads to a bias in the estimate of the luminosity function).
We use the coherent method of Avni \& Bahcall (1980) to maximise the
information in the combined samples.
In calculating the absolute magnitudes of the quasars we use K-corrections as defined
by Schmidt \& Green (1983), i.e. assuming a featureless power-law slope with a
spectral index $\alpha=-0.5$. We compared this approximation to 
the K-corrections tabulated by Cristiani \& Vio (1990) 
and found that there was no significant difference in the
estimated luminosity function due to the difference in K-correction 
within the redshift range used in this paper.

The above method has been used to construct the differential luminosity function
for the surveys (Fig. 1). The redshift slices have been chosen
to be the same as those used in BSP. The first impression is that
the luminosity function changes shape as a function of redshift, in direct
contrast to the PLE model, and that the luminosity function in the 
lowest redshift slice looks like a featureless power-law with no break at all
for $M_{B} \le -23$.  We investigate the evolving shape of the luminosity
function in the following sections.  We assume the Hubble
constant H$_0 = 50$\,km\,s$^{-1}$\,Mpc$^{-1}$, zero cosmological
constant, and the deceleration parameter $q_0 = 0.5$ unless otherwise stated.

\begin{figure}
\epsfig{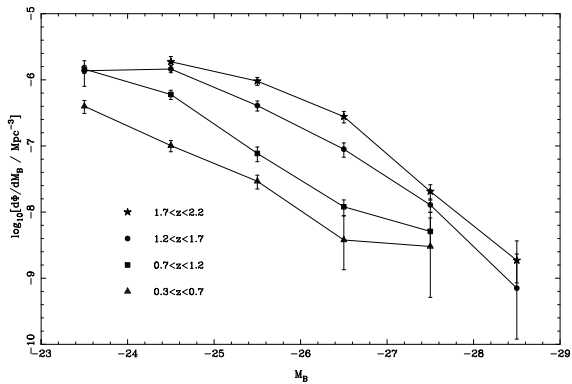}
\caption{The differential luminosity function using the Edinburgh,
AAT, MBQS and SA94 surveys, for $q_0,h = 0.5$.}
\label{fig1}
\end{figure}

\section{COMPARISON WITH PURE LUMINOSITY EVOLUTION MODELS}

In order to assess whether the observed luminosity function agrees with the PLE model we
carry out two tests. The first is a non-parametric test, comparing the observed
cumulative distribution in luminosity with that predicted by PLE using the
one-dimensional one-sample Kolmogorov-Smirnov test (e.g. Conover 1980) to test the 
null hypothesis that both the observed and model distributions are drawn 
from the same parent population.

\subsection{Comparison with the standard model}

We bin the data into redshift slices as above and calculate the observed
cumulative distribution in absolute magnitude (note that we cannot simply
compare observed and predicted cumulative luminosity functions as the KS test 
requires that each data point have equal weight, which is not true for the 
volume-weighted luminosity function).
The theoretical cumulative distribution is calculated from the PLE model of
BSP for each object with absolute magnitude $M_{B}$ and redshift $z$;
\begin{equation}
	N(<M_{B}) = \int_{z_{1}}^{z_{2}} \int_{M_{B-bright}}^{M_{B}} 
		d\phi(M_{B}^{'},z)\Omega(M_{B}^{'},z) \frac{dV}{dz} 
                dM_{B}^{'} dz \\
\end{equation}
where the redshift limits $z_{1}, z_2$ are determined by both the limits of the
redshift slices and the distance limits for detection given the
apparent magnitude limits (both bright and faint) of the survey, and 
$M_{B-bright}$ is the luminosity corresponding to the bright flux limit at
$z$.
$\phi(M_{B}^{'},z)$ is determined from the model parameters, and
$\Omega(M_{B}^{'},z)$ is the effective area searched to find that object.

We use three different subsets of the combined Edinburgh and AAT surveys;\\
(A) The whole of both surveys. \\
(B) The data from both surveys brighter than the absolute magnitude
corresponding to the BSP break in each redshift slice. This absolute magnitude
was calculated from equation (2) with $z$ taken to be the upper redshift 
limit of each redshift slice.
Under the null hypothesis that we are testing, using this method should mean
that the data used has the same shape distribution, regardless of redshift, if
PLE is an adequate description of the data. \\
(C) The whole of the Edinburgh survey alone.

\begin{figure}
\epsfig{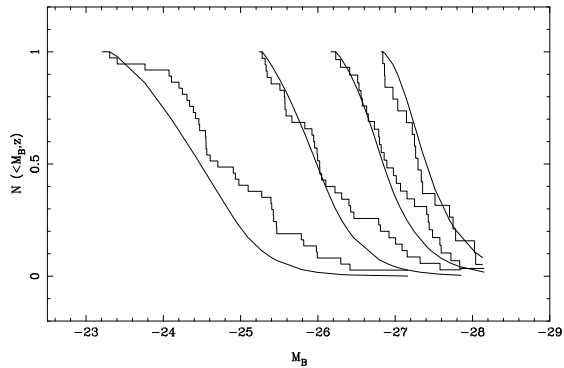}
\caption{The observed cumulative distributions in absolute magnitude 
from the Edinburgh survey and the predicted distributions from the PLE 
model of BSP}
\label{fig2}
\end{figure}

Figure 2 shows the observed and model distributions for case C, i.e. for the
Edinburgh survey alone.
Table 1 shows the probabilities that the null hypothesis is true
in each redshift slice for each of the three cases outlined above.
In cases B and C the probability that the
model describes the data in the lowest redshift slice is unacceptable at
a significance level of $0.1\%$.  
We also find that if we carry out a two-dimensional KS test (Peacock 1983) just
on the Edinburgh sample over the entire range of redshifts, 
we obtain a probability of $2\%$ that the PLE model describes the data adequately.  
This rejection of the PLE model is not
found when testing case A, a reflection of the fact that it is the
high luminosity quasars which are responsible for the effect.  This is
not surprising:  the PLE model was developed to fit the fainter AAT data,
which we continue to use in this analysis, plus the brighter data of Schmidt
\& Green (1983), which we have previously argued is significantly incomplete
(Goldschmidt {\em et al.}, 1992)
and which we have replaced by the Edinburgh quasar survey.  We should therefore
expect to see the most significant differences between the model and the Edinburgh data.

\begin{table}
\caption{The KS probabilities of the three subsets of the data 
defined in the text being consistent with the PLE model 
of BSP}
\begin{tabular}{|clll|}
\hline
redshift            & case A & case B  & case C  \\
\hline
$0.3 \le z \le 0.7$ & $0.10$ & $0.001$ & $0.001$ \\
$0.7 \le z \le 1.2$ & $0.78$ & $0.17$  & $0.21$ \\
$1.2 \le z \le 1.7$ & $0.95$ & $0.42$  & $0.27$ \\
$1.7 \le z \le 2.2$ & $0.38$ & $0.65$  & $0.80$ \\
\hline
\end{tabular}
\end{table}

\subsection{Comparison with general power-law models}

We can extend our analysis to test whether the data can be fitted by
{\em any} evolving power-law model in which the power-law 
index remains constant. 
Both the PLE model, at magnitudes more luminous than the BSP break, and the
Hawkins \& V\'{e}ron model are examples of this class of model.  
In this section we fit a single power-law model to the data in each redshift slice,
but only at luminosities higher than the BSP break luminosity.  We then
test the null hypothesis that these bright-end power-law indices in each redshift slice 
have the same value.   

The best-fit values for the indices are
calculated using maximum likelihood assuming a single power-law fit to the
data more luminous than the BSP break in each redshift slice,
\begin{equation}
d\phi(M_{B})=\phi^{*}10^{0.4(\hat{\alpha} -1) M_{B}} dM_{B}
\end{equation}
where $\hat{\alpha}$ is the estimate of the index of the power law.
The faintest absolute magnitude in each redshift
slice used to fit the model to the data is calculated by taking a comoving space
density $\rho=10^{-6.4} {\rm Mpc}^{-3}$ ($q_0 = 0.5$) 
and finding the corresponding absolute magnitude in the BSP model at different redshifts.
Under the null hypothesis this should give a constant index for all redshifts.
The answer should not be overly dependent on the value of $\rho$ chosen,
although too
low a value will result in too little relevant data being used to fit the
model, thereby reducing the statistical significance. Too high a value will result in
some of the data from the flat part of the luminosity function being used, 
again underestimating the true significance.

Errors on $\hat{\alpha}$ are calculated by assuming a $\chi^{2}$ distribution for
$S-S_{max}= -2 log(L/L_{max})$ where $L$ is the likelihood function.
For $q_{0}=0.5$ the best-fit power-law index increases 
from $\hat{\alpha} = 2.7$ in the lowest
redshift slice to $\hat{\alpha} = 4.1$ in the highest slice (Fig.\,3).
A single value for $\hat{\alpha}$ is ruled out at a
significance level of $0.1 \%$. For $q_{0}=0.1$ the index increases from
$\hat{\alpha}=2.6$ to $\hat{\alpha}=3.6$ and a single value for
$\hat{\alpha}$ is unacceptable at a significance level of $2\%$.

\begin{figure}
\epsfig{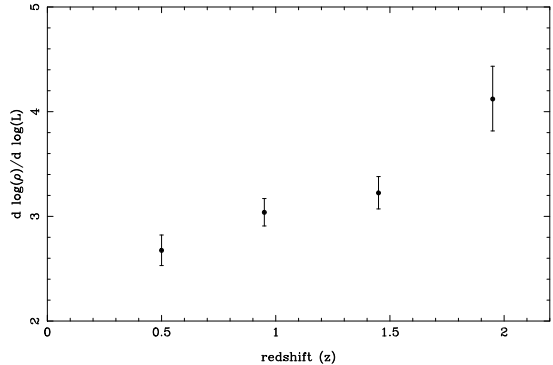}
\caption{Maximum likelihood estimates of the power-law index 
of the luminosity function at absolute magnitudes more luminous than 
the BSP break, for $q_0 = 0.5$.}
\label{fig3}
\end{figure}

The evidence presented in this section shows clearly that the high-luminosity 
part of the luminosity function does {\em not} evolve according to the 
expectations of pure luminosity evolution.  The slope of the luminosity function
at $M_B \la -26$ displays significant steepening with redshift.

\section{THE QUASAR LUMINOSITY FUNCTION AT LOW AND HIGH REDSHIFTS}

The analysis in the previous section showed that the power-law index
changed with redshift for quasars more luminous than the break.  We have
also previously remarked that there appears to be no evidence for a break
in the luminosity function at low redshift.   
In this section we fit a single power-law model to {\em all} 
the quasars brighter than $M_{B}=-23$
in the lowest redshift bin with $0.3 \le z \le 0.7$;
\begin{equation}
d\phi(M_{B},z) = \phi^{*}(1+z)^{\gamma}10^{-0.4(\alpha+1)M_{B}}dM_{B}dz
\end{equation}
using maximum likelihood as above. The best fit to these parameters 
are $\alpha=-2.6\pm 0.2$ and $\gamma=9.1\pm 2.0$. 

To test the goodness-of-fit of the single power-law we could use a binned chi-squared
test.  This is not the most efficient way of testing the model as chi-squared 
cannot cope with incomplete bins. We prefer instead to use a Kolmogorov-Smirnov
test to estimate goodness-of-fit, following previous work such as Boyle et al.
(1988 \& 1991).  This procedure suffers from the problem
that the slope of the power-law is a free parameter in the model, and the value
of the slope has been found by fitting to the data.  Thus the significance level at
which the model can be rejected is actually an overestimate:  strictly speaking, 
in this application the KS test can only rule out models, but the fact that
there is a value for a single power-law model which does fit should indicate
to us that there is no justification for the pursuit of a more complicated model
at low redshifts.  This is indeed the result we obtain.
The KS test shows that this model is acceptable, with a significance level for
rejection of $17\%$.

Conversely, at redshifts $1.7 < z < 2.2$ it is clear 
that the luminosity function cannot be parameterised as a single power-law.
The data can be fitted by either a dual
power-law model as described previously
or a variety of other functional forms. For example, we can fit
a Schechter function with uniform luminosity evolution to the data at
$1.7 \le z \le 2.2$, where the model is parameterised as;
\begin{eqnarray}
& & \frac{d\phi}{dM}_{B}(M_{B},z) = \nonumber \\
& & \phi^{*}10^{-0.4(\alpha+1)(M_{B}-M^{*}(z))}
\exp[-10^{-0.4(M_{B}-M^{*}(z))}]
\end{eqnarray}
where $M^{*}(z)=M_{0}-2.5\gamma\log10(1+z)$.
The best-fit values of the parameters are $\alpha=-1.7\pm 0.3$, 
$M_{0}=-23.7\pm 0.4$ and $\gamma=2.5\pm0.3$. The fit is acceptable,
with a significance level for rejection of $10\%$.
The single power-law model which fits the data at low redshifts (as described
above) can be rejected at a significance level $<0.1\%$ in the redshift 
range $1.7 \le z \le 2.2$.

\section{DISCUSSION}

From the above analysis we reach the following conclusions.
\begin{enumerate}
\item The shape of the quasar luminosity function changes shape with
redshift, in a manner that cannot be described as pure luminosity evolution.
Specifically, the slope of the luminosity function at high luminosities is
significantly steeper at redshifts $z \sim 2$  than it is at $z \sim 0.5$.
\item At redshifts $z \la 1$ the luminosity function may be described by a single
power-law, and there is no evidence for any feature in the luminosity function that 
may be used as a tracer of luminosity evolution.  Conversely, at $z > 1$ the
luminosity function cannot be described by a single power-law, as previously 
found by BSP, and there is a break in the luminosity function at $M_B \sim -26.5$.
The luminosity function at high redshift may be described by a number of functional
forms such as the two-power law model of BSP or indeed by a Schechter function.
\end{enumerate}

The consequence of
these conclusions are that it cannot be shown that the quasar population
experiences luminosity evolution.  It may be that quasars are long-lived
and that they do indeed dim with cosmic epoch in a luminosity-dependent manner
so as to produce the observed evolution.  But it is equally possible that 
quasars are short-lived phenomena and that the observed evolution is a more
complex mixture of luminosity-dependent density evolution.  In fact, recent
models (Goldschmidt 1993, Percival, Miller \& Goldschmidt 1997) 
suggest that such evolution might be expected if quasars are
short-lived symptoms of galaxy mergers in a CDM-type (``bottom-up'')
universe.  In this case, it is only possible to make progress in understanding
quasar evolution  by constructing a specific model such as the one just described
and then comparing the predictions of the model with the observed luminosity function.
It is not possible to deduce model-independent conclusions about whether or not
quasars undergo luminosity evolution from consideration of the
observed luminosity function alone.

One piece of information which must be a powerful clue to the type of model that is 
required, however, is the observation that the amount of density evolution is
greatest at intermediate luminosities ($M_B \sim -26$), and appears to be less at
higher quasar luminosities.  Extrapolation of this result would indicate that
at $M_B \sim -29$ the comoving space density of quasars may have remained roughly
unchanged since $z = 2$!  The existing data are too noisy at high luminosities
and low redshifts to demonstrate this unambiguously, and we must await larger-area
surveys to provide better statistical evidence for the most luminous quasars.

\vspace*{0.5 true in}\noindent
{\large\bf Acknowledgements}\\

Data reduction and analysis were carried out on STARLINK. 
P. Goldschmidt acknowledges support from PPARC.

\end{document}